\documentclass[12pt]{article}
\usepackage{graphicx}
\usepackage{amsmath}
\usepackage{amssymb}
\usepackage{hyperref}
\numberwithin{equation}{section}

\def\bC{\mathbb{C}}
\def\bH{\mathbb{H}}
\def\bR{\mathbb{R}}

\def\cA{\mathcal{A}}

\def\IC{\mathbb{C}}

\def\cM{\mathcal{M}}
\def\cN{\mathcal{N}}
\def\cO{\mathcal{O}}

\def\fg{{\displaystyle\mathfrak{g}}}

\def\SU{\mathrm{SU}}
\def\su{\mathfrak{su}}
\def\SO{\mathrm{SO}}

\def\Sp{\mathrm{Sp}}
\def\U{\mathrm{U}}

\def\diag{\mathop{\mathrm{diag}}}

\def\tr{\mathrm{tr}}

\def\hkq{/\!/\!/}

\def\hk{hyperk\"ahler}
\def\CM{{\cal M}}

\begin{document}

\begin{titlepage}

\begin{flushright}
IPMU-11-0171
\end{flushright}
\vskip 3cm

\begin{center}
{\Large \bfseries
On 6d $\cN=(2,0)$ theory
compactified \\[.6em]
on a Riemann surface with finite area
}

\vskip 1.2cm

Davide Gaiotto$^1$,
Gregory W. Moore$^2$ and Yuji Tachikawa$^3$

\bigskip

$^1$ School of Natural Sciences, Institute for Advanced Study, \\
Princeton, NJ 08504, USA

\medskip

$^2$ NHETC and Department of Physics and Astronomy, Rutgers University,\\
 Piscataway, NJ 08855, USA

\medskip

$^3$ IPMU, University of Tokyo, Kashiwa, Chiba 277-8583, Japan

\vskip 1.5cm

\textbf{abstract}
\end{center}

\medskip

We study 6d $\cN=(2,0)$ theory of type $\SU(N)$ compactified on Riemann surfaces with finite area, including spheres with fewer than three punctures. The Higgs branch, whose metric is inversely proportional to the total area of the Riemann surface, is discussed in detail.
We  show that the zero-area limit, which gives us a genuine 4d theory, can involve a Wigner-\.{I}n\"on\"u contraction of   global symmetries of the six-dimensional theory. We show how this explains why   subgroups of $SU(N)$ can appear as the gauge group in the 4d limit.
As a by-product we suggest that half-BPS codimension-two defects in the six-dimensional $(2,0)$ theory have an operator product expansion whose operator product coefficients are four-dimensional field theories.

\bigskip
\vfill
\end{titlepage}

\tableofcontents

\section{Introduction}

In the past few years we have learned many things about
a class of four dimensional field theories - sometimes called
``theories of class $S$'' -  obtained by compactifying
the six-dimensional $\cN = (2,0)$ theory on a Riemann surface $C$.
This note discusses one subtlety which can arise when deriving the
four-dimensional theory from the six-dimensional theory.
In the process
we clarify some aspects of the behavior of the four-dimensional theories in
weak-coupling limits defined by degenerations of the complex structure
of $C$. Our considerations naturally suggest the existence of
an ``operator product expansion'' (OPE) of codimension two supersymmetric
defects in six-dimensional $(2,0)$ theory
 whose OPE coefficients are four-dimensional field theories.
 Our discussion will be
somewhat informal and makes no pretense to being fully systematic
or complete.

To be more precise, we will focus on the  $A_{N-1}$ theories of class $S$.
This means we begin with the six-dimension $\cN = (2,0)$ theory of type%
\footnote{In order to keep this paper brief  we will not
be extremely careful about the precise global form of the gauge group. }
$\SU(N)$ on $\bR^{1,3} \times C$ where $C$ is a
punctured Riemann surface of genus $g$. The theory is partially topologically twisted
in order to preserve $d=4, \cN =2$ supersymmetry and at each puncture $p_i$ there are
certain half-BPS  codimension-two defects $D(\rho_i)$, where $\rho_i$ is a homomorphism
$\rho_i: \SU(2) \to \SU(N)$. This construction goes back to \cite{Klemm:1996bj,Witten:1997sc}
and its study was rekindled in  \cite{Gaiotto:2009we,Gaiotto:2009hg}, to which
we refer for more details. The associated  four-dimensional theory at
scales much larger than those of $C$ is denoted   $S_N[C,D]$ where $D$ stands for the collection $\{D(\rho_i)\}$.
For certain choices of $C$ and $D(\rho_i)$ there can be difficulties in taking
the   four-dimensional limit.

In this paper we illustrate the above-mentioned difficulties by focusing on the
Higgs branch of $A_{N-1}$ theories of class $S$ when the area of $C$ is nonzero.%
\footnote{Note that the Coulomb branch only depends on the complex structure of $C$,
and is independent of the area.
The area introduces a mass scale, thus breaking the superconformal symmetry.
However, the system still has the $\SO(3)_R\times \U(1)_R$ symmetry,
which is the unbroken part of the original $\SO(5)_R$ symmetry of the 6d $\cN=(2,0)$ theory.
In terms of the 't Hooft anomaly coefficients involving these R-symmetries and gravity, we can still define two central charges $a$ and $c$, or equivalently $n_v$ and $n_h$. These equal the standard central charges defined in terms of energy-momentum tensors when the limit $\cA\to 0$ can be naively taken.}
As we show in Sec. \ref{asCoulomb}
below, the \hk\  metric on the Higgs
branch only depends on the metric on $C$ through the total area $\cA$,
a result which is  in harmony with the nice recent discussion of \cite{Anderson:2011cz}.
Thus, the limit we focus on is $\cA\to 0$.
The dependence of the Higgs branch on the area is simple:
\begin{equation}\label{eq:A-dep}
ds^2_{\cA} = \cA^{-1} ds^2_{\cA = 1} .
\end{equation}
Evidently, the limit $\cA \to 0$ does not make sense without
some further discussion. If we fix a point
on the Higgs branch then the limit can be taken by simultaneously
restricting attention to fields which lie at a finite distance from that chosen point.
Now, a generic point on the Higgs branch breaks R-symmetries and global symmetries.
The absence of a point preserving UV R-symmetries is an indication that
the IR limit might contain very different physics from what would naively
expect. The situation is very similar to the trichotomy between
good/bad/ugly 3d gauge theories discussed in \cite{Gaiotto:2008ak} and
in fact in   Sec. \ref{asCoulomb} we relate our discussion directly to
that work. In the good theories, there is a region of the Higgs branch
which looks like a cone. Choosing the vacuum at the tip of the cone,
none of the expected R-symmetries or global symmetries are broken in
the $\cA \to 0$ limit. In the ugly theories, there is still a natural vacuum which does not break the expected symmetries,
but it is not a conical singularity (or possibly is locally the product of a smooth part and a conical singularity).
Thus free hypermultiplets appear in the IR.
In the bad theories, there is no point on the moduli space which
preserves the symmetries: We need further input in order to understand
the IR physics.

The above subtleties of the  $\cA \to 0$ limit are closely related to   the behavior
of $S_N[C,D]$ when the complex structure on $C$ degenerates. As first
stressed in \cite{Gaiotto:2009we} this behavior is related to the gauging of
global symmetries of theories of class $S$. Let us recall the
basic assertion. Consider a
separating degeneration where $C$ splits  into a one-point union of
$C_L$ and $C_R$ at a common point $p$. The degeneration splits  the set of
defects into $D_L$ and $D_R$. A neighborhood of this point, in the
moduli space of complex structures on $C$, can be parametrized by
introducing coordinates $z_L, z_R$ near $p_L\in C_L$ and $p_R\in C_R$ and
sewing the surfaces together using
the plumbing fixture $z_L z_R = q$. The sewn surface near the degeneration
limit is denoted $C_L \times_q C_R$ and the degeneration limit is  $q\to 0$.
Then the basic gluing law states that:
\begin{equation}\label{eq:glue-gauge}
S_N[C_L \times_q C_R, D_L \cup D_R] = S_N[C_L, D_L \cup D_f] \times_{SU(N),q} S_N[C_R, D_R \cup D_f]
\end{equation}
where on the right-hand side $D_f$ refers to the so-called ``full puncture''
with full $\SU(N)$ global symmetry and $\times_{SU(N),q}$ means that the diagonal
subgroup of the global $\SU(N)\times \SU(N)$ global symmetry of the two full punctures
is gauged with the coupling constant $q\sim e^{i \pi \tau}$. It was shown in
\cite{Moore:2011ee} that \eqref{eq:glue-gauge} naturally leads to a notion of a ``two-dimensional
conformal field theory valued in four-dimensional field theories,''
a notion which has yet to be made completely precise.  Unfortunately,
there are certain cases of \eqref{eq:glue-gauge} which are not strictly
true. In these cases  the statement must be amended.
In particular, there are cases when only a subgroup
of the diagonal $\SU(N)$ gauge group is gauged.
 This was already noted in \cite{Gaiotto:2009we}
and was discussed further in \cite{Chacaltana:2010ks,Chacaltana:2011ze}; even the prototypical example of Argyres and Seiberg \cite{Argyres:2007cn} involved the subgroup $\SU(2)$ of $\SU(3)$.
The subtlety appears when one or both halves $C_L$, $C_R$ are spheres with certain combinations of punctures $D(\rho_i)$ which are ``too small''.
In the present paper we give a complementary discussion of the subtleties.

In a nutshell, we find that even when the combination of $D(\rho_i)$ is not good, the theory at finite $\cA$ always has $\SU(N)$ flavor symmetry associated with the defects at $p_L$ and $p_R$. We will see, however, that at no point in the vacuum moduli space is all of $\SU(N)$ preserved;  at most a subgroup $H\subset \SU(N)$ remains unbroken. Then in the $\cA\to 0$ limit, the broken part of $\SU(N)$ is contracted \`a la \.In\"on\"u-Wigner, and cannot  even be seen acting on the theory in the four-dimensional limit. Instead, in papers \cite{Chacaltana:2010ks,Chacaltana:2011ze} the authors identified   $H$ using  various indirect means.
We will introduce the notion of fusion, or OPE, of two or more defects, which captures the subtleties of the $\cA \to 0$ limit. Note that although in two-dimensional rational conformal field theories
one can always represent the OPE of two vertex operators as the sewing in of a trinion into the surface, this is not the case in the most general non-rational conformal field theories,
in particular Toda theories. In general two semi-degenerate representations of Toda have an OPE which consists of an integral over
some other class of semi-degenerate representations in the intermediate channel, and cannot be produced by a straightforward sewing procedure: the sewing
would produce an integral over non-degenerate representations.

The rest of the paper is organized as follows. In Sec.~\ref{aperitif}, we consider two easy cases, namely 6d theory on a torus and on a sphere with two full punctures,
 to see the area dependence explicitly and observe two different behaviors in the $\cA\to0$ limit.
In Sec.~\ref{higgs}, we study the dependence of the Higgs branch of the system on the metric of $C$ from various perspectives. We learn that the Higgs branch only depends on the total area of $C$, we discuss the basic
 trichotomy for the behavior in the $\cA\to 0$ limit, and devise a method to obtain the Higgs branch as the hyperk\"ahler quotient constructed out of a few basic ingredients. We also study a general way to deform the metric of a hyperk\"ahler manifold with a group action.
In Sec.~\ref{4d}, we apply the knowledge obtained to the analysis of 4d theories.
We return to the exceptions to the gluing law \eqref{eq:glue-gauge}. We will gain more insight, for example, as to how an $\SU(2)$ gauge group can arise in the strong-coupling dual to the $\SU(3)$ gauge theory with six flavors. This is one of the cases where
the factorization statement \eqref{eq:glue-gauge} must be amended. The considerations of
 factorization naturally lead one to the
study of the behavior of two half-BPS defects of type $D(\rho)$ when
they are close together. We believe there should be an analog of the
operator product expansion whose coefficients are four-dimensional
field theories. We briefly introduce that idea in Sec.~\ref{ope} below.
%

\section{Two easy pieces}\label{aperitif}
\subsection{Torus}
Consider 6d theory of type $A_{N-1}$ on a rectangular $T^2$, with lengths of sides given by $R_5$ and $R_6$. Its moduli space is the same as that of the 5d maximally-supersymmetric $\SU(N)$ Yang-Mills, with coupling constant $1/g^2_{5d} \sim 1/R_6$, compactified on a circle with circumference $R_5$.
Up to the identification by the Weyl group,
the five scalars give $(\bR^5)^{N-1}$, and the Wilson line around $S^1$ gives $(S^1)^{N-1}$.
If we view this $\cN=4$ theory as an $\cN=2$ theory the moduli space contains
the Coulomb branch $(\mathfrak{t}\oplus\mathfrak{t})/W \cong (\mathfrak{t}\otimes \bC)/W$
and the Higgs branch $(\mathfrak{t}\oplus\mathfrak{t}\oplus\mathfrak{t}\otimes T)/W
\cong (\mathfrak{t}\otimes \bH)/\widehat{W}$, where $\mathfrak{t}$ and $T$ are the
Cartan subalgebra and the Cartan subgroup of $\SU(N)$, $W\cong \mathfrak{S}_N$ is the Weyl
group and $\widehat{W}$ is the affine Weyl group.

Let the periodicity of the scalars parameterizing $T$  be $2\pi$, which means we set \begin{equation}
R_5A_5=\diag(\phi_1,\ldots,\phi_N)
\end{equation} with the identification $\phi_i\sim \phi_i+2\pi$. Then the kinetic term of $\phi_i$ is given by \begin{equation}
\sim \int dx_5 \frac{1}{g^2_{5d}}\tr (\partial_\mu A_5)^2 \sim \frac{1}{R_5R_6} \sum_i \partial_\mu\phi_i\partial_\mu\phi_i.
\end{equation} Therefore, the metric of the Higgs branch has the area dependence of the form $ds^2 = (\cA)^{-1}ds_{\cA=1}^2$.
  As discussed in the Introduction, we must choose a point around which to take
  the $\cA \to 0$ limit.
  If we choose the origin of the Higgs branch, which is an orbifold point,   the
  limit   $\cA\to 0$ turns the Higgs branch into its ``tangent space''
   $(\mathfrak{t}\oplus\mathfrak{t}\oplus\mathfrak{t}\oplus\mathfrak{t} )/W$, which is the Higgs branch of 4d $\cN=4$ super Yang-Mills. Note that the topology of the Higgs branch has changed.

\subsection{Sphere with two full punctures}\label{T*G}
\begin{figure}
\[
\includegraphics[width=.4\textwidth]{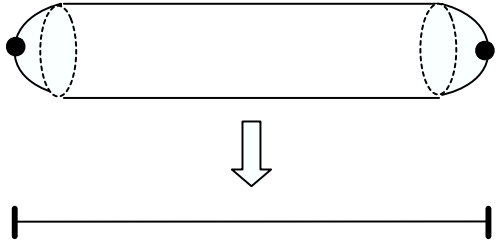}
\]
\caption{6d theory on a sphere with two full punctures, and its reduction to 5d theory. The  punctures become boundary conditions.\label{fullcigar}}
\end{figure}

Next, let us consider 6d theory of type $A_{N-1}$ on a sphere of area $\cA$, with two full punctures, each carrying $\SU(N)$ global symmetry.
We choose the metric on the sphere so that it looks like a cylinder of circumference $R_6$ and length $R_5$ with $R_6\ll R_5$, capped by disks each with a full puncture at the center, see Fig.~\ref{fullcigar}.

This system can be analyzed as the 5d maximally-supersymmetric $\SU(N)$ Yang-Mills with coupling constant $1/g^2_{5d}\sim 1/R_6$, put on a segment with length $R_5$, with Dirichlet boundary condition at both ends. The BPS equation whose solution corresponds to a point in the Higgs branch is the Nahm equation on the segment $s\in [0,R_5]$:
\begin{equation}\label{eq:NahmEq}
\frac{d}{ds}\Phi_i + [A_s,\Phi_i]= \epsilon_{ijk} [\Phi_j,\Phi_k]
\end{equation}
where $i,j,k$ run from $1$ to $3$. The Dirichlet boundary conditions of the
Yang-Mills theory imply that $\Phi_i(s)$ should be regular at both boundaries.
We identify two solutions
related by a gauge transformation $h:[0,R_5] \to \SU(N)$ such that $h(0)=h(R_5)=1$.
The metric on the moduli space comes from the kinetic terms in the 5d Lagrangian,
and, analogously to the case in the previous subsection, it has a factor of $1/\cA \sim 1/(R_5R_6)$ in it.
We will denote   this hyperk\"ahler moduli space by $I(\cA)$.
Since the group of \emph{all} maps $h:[0,R_5] \to \SU(N)$   acts on solutions to \eqref{eq:NahmEq} there is a global
$\SU(N) \times \SU(N)$ symmetry acting on $I(\cA)$,  where the two factors are obtained from $h(0)$ and $h(R_5)$.

The moduli space $I(\cA)$ can be parametrized by $g=P\exp \int_0^{R_5} A_s ds$ and $\Phi_i(0)$. Therefore it is topologically $\simeq \SU(N)\times \su(N)^3$.
Let us consider a point $(g,\phi_i)$ in it. Then the global symmetry element $(h_1,h_2)\in \SU(N)_1\times \SU(N)_2$ acts via \begin{equation}
(g,\phi_i)\to (h_1 gh_2^{-1},h_1 \phi_i h_1^{-1}).
\end{equation}
Of course, at a general point on the moduli space the global symmetry is
broken to a discrete group (the center of $\SU(N)$, diagonally embedded).
However,  even when $\phi_i=0$, the global symmetry $\SU(N)_1\times \SU(N)_2$ is spontaneously broken to a diagonal subgroup $\SU(N)$ specified by $h_1=gh_2g^{-1}$. In particular, there is no point where the whole of the global symmetry $\SU(N)^2$ is unbroken. The largest
isotropy group of any point is $\SU(N)$.

Now let us consider taking the limit $\cA\to 0$. Once again, as discussed
in the Introduction,  one must
choose a point around which to expand. It is instructive to see how the
global symmetries behave in this limit. The most symmetric point we can choose is
 $(1,\vec 0) \in \SU(N)\times \su(N)^3$. As before,
    the limiting metric is just the flat metric on the
    tangent space at  $(1,\vec 0)$, which is isomorphic to $\su(N) \oplus  \su(N)^3 \cong \su(N)\otimes \bH$.
    The $\SU(N)\times \SU(N)$ symmetry was broken to the diagonal $\SU(N)$. The broken anti-diagonal
    symmetries contract to translations by  $\bR^{N^2-1}$. The isometry group of the
    IR limit, commuting with the hyperk\"ahler structure, is just the semidirect product of
    translations with   $\Sp(N^2-1)$.

This example illustrates two points: i) some of the global symmetry at nonzero $\cA$ can get contracted
in the limit, and ii) the global symmetry after $\cA\to 0$ limit can be enhanced.

\section{Higgs branch for $C$ with finite area}\label{higgs}
Having seen two easy examples, let us discuss the general case
of $A_{N-1}$ theories of class $S$ described in the Introduction.
As we mentioned there, each defect is  labeled by a map $\rho:\SU(2)\to \SU(N)$.
(These defects admit mass deformations. However in this paper we take the
mass deformations to be zero.)
Equivalently, $\rho$ is given by a partition
$(\lambda_1,\lambda_2,\ldots)$ of $N$. We use the notation so that, for example,
 $\rho=[3^21^2]$ stands for the partition $8=3+3+1+1$.
The puncture of type $\rho$ has a flavor symmetry $G^\rho$, which is the commutant of the image of $\rho$ inside $G$.
The punctures corresponding to $f=[1^N]$ and $s=[N-1,1]$ are
particularly important and are called \emph{full} and \emph{simple}, respectively; the puncture $\rho=[N]$ corresponds to the absence of the puncture altogether. The full puncture has $\SU(N)$ flavor symmetry, and the simple puncture has $\U(1)$ flavor symmetry.
We first analyze the Higgs branch of this  system in two ways in Sections \ref{asCoulomb}
and \ref{subsec:asQuotient}, then we apply those two viewpoints.

\subsection{As the Coulomb branch of 5d super Yang-Mills on $C$}\label{asCoulomb}
Let us consider our 4d system on $S^1$, of circumference $R$.
It has 3d $\cN=4$ symmetry. The Higgs branch does not depend on $R$; but as the metric of the moduli space has mass dimension two and one in spacetime dimension four and three, respectively, it is natural to set \begin{equation}
 ds^2(\text{4d Higgs branch})=  R^{-1} ds^2(\text{3d Higgs branch}) . \label{trivial43}
\end{equation}

We now have 6d theory compactified on $S^1\times C$. We can perform the compactification on $S^1$ first, and regard the system as
5d maximally-supersymmetric $\SU(N)$ Yang-Mills on $C$ with codimension-two defects $D(\rho_i)$.
The coupling constant is as always $1/g^2_{5d}\sim 1/R$. Since 5d SYM is IR free we
can identify the defects as 3d superconformal field theories coupled to the bulk.
It turns out these are just the theories   called $T_\rho[\SU(N)]$   in \cite{Gaiotto:2008ak}.
This procedure is effectively the 3d mirror operation, and as such the original Higgs branch is
the Coulomb branch of this 3d system obtained by compactifying the 5d SYM on $C$.

As a 3d theory, our 5d SYM on $C$  has an infinite-dimensional gauge group of  maps from $C$ to $\SU(N)$.
The 5d kinetic term \begin{equation}
\int_{\bR^3} d^3x \int_C dzd\bar z e^\phi \frac{1}{g^2_{5d}}\tr F_{\mu\nu}F_{\mu\nu}
\end{equation}
can be thought of defining a coupling matrix on the gauge algebra of maps  $X,Y:C \to \su(N)$ via \begin{equation}
\int_C dzd\bar z e^{\phi} \frac{1}{g^2_{5d}} \tr XY.
\end{equation}
Here we used the complex structure and the Weyl mode to express the 2d metric on $C$.
This infinite-dimensional group is always broken down to $\SU(N)$, which corresponds to constant maps from $C$ to $\SU(N)$.
Effectively, our 3d theory is just $\SU(N)$ $\cN=4$ theory coupled to $g$ hypermultiplets in the adjoint representation of $\SU(N)$ and $T_{\rho_i}[\SU(N)]$ where $i=1,\ldots,n$ and $g$ is the genus of $C$. The adjoint hypermultiplets come from the zero modes of   $A_z$, $A_{\bar z}$ on $C$.

The metric of the Coulomb branch only depends on the coupling constant of the unbroken gauge group, and not on the coupling matrix of the broken part of the gauge fields.
The coupling constant of the unbroken $\SU(N)$ gauge field is given by \begin{equation}
\frac{1}{g_3^2}=\frac{\int_C dz d\bar z e^{\phi} }{g_5^2} = \frac{\cA} R.\label{3dcoupling}
\end{equation}
As this is the only scale in the system, the metric on the 3d Coulomb branch has an overall factor of $R/\cA$. Combining with \eqref{trivial43}, we see that the 4d Higgs branch has an overall factor of $1/\cA$.
Let us stress that the metric does not depend on the detailed form of the Weyl mode $e^\phi$.

Recall that $T_\rho[\SU(N)]$ has a linear quiver realization \cite{Gaiotto:2008ak}: for a partition $\rho=[\lambda_1,\lambda_2,\ldots,\lambda_k]$ with $\lambda_1 \ge \lambda_2\ge \cdots$, the quiver is
 \begin{equation}
\underline{\SU(N)} - \U(n_1)-\U(n_2)-\cdots -\U(n_{k-1})
\end{equation}
where $n_s=\sum_{s<t} \lambda_t$; the underlined group is a flavor symmetry.
$T_\rho[\SU(N)]$ is defined to be the limit where the gauge coupling of all the gauge groups are taken to infinity.

Then our Coulomb branch is obtained by taking the linear quiver realizations of $T_{\rho_i}[\SU(N)]$ for each $\rho_i$, and coupling it to an $\SU(N)$ and $g$ adjoint hypermultiplets \cite{Benini:2010uu}. We keep the gauge coupling of the central $\SU(N)$ finite,
 given by \eqref{3dcoupling}, but take the coupling constants of all the other gauge groups to be infinitely large.

Let us consider the genus zero case, and consider defects labeled by partitions $\rho_i=[\lambda_{i,1},\lambda_{i,2},\ldots]$. Then  the central $\SU(N)$ has in total
\begin{equation}
N_f=\sum_i  n_{i,1} = \sum_i (N-\lambda_{i,1})
\end{equation}
 fundamental flavors.
  Depending on whether $N_f\ge 2N$, $N_f=2N-1$, or $N_f\le 2N-2$, the dynamics of the $\SU(N)$ gauge multiplet is ``good'', ``ugly'' or ``bad'' in the terminology of \cite{Gaiotto:2008ak}. In our context, when it is good the $\cA\to 0$ limit gives us an interacting 4d theory;
when it is ugly the $\cA\to 0$ limit gives us a free 4d theory, or an interacting theory with a free subsector; when it is bad, more data is needed to specify an $\cA \to 0$ limit.
In contrast to the example   in Sec.~\ref{T*G} there is    no canonical place in the moduli space to take the limit. In the ``bad''
cases the $R$-symmetries and global symmetries in the UV and IR theories can be quite different.
When $g>1$, the theory is always good. When $g=1$, the theory is bad when there is no puncture, ugly when there is only one simple puncture, and good otherwise.

Let us conclude with several remarks:

\begin{enumerate}

\item
This approach to the moduli space tells us when the limit $\cA\to 0$ is easily taken.
But it does not give us a way to calculate the metric, because we do not quite know how to determine the exact, quantum-corrected metric on the Coulomb branch of a 3d $\cN=4$ gauge theory yet. However, this expression has the virtue of showing its independence 
from the nonzero modes of the Weyl factor of the metric on $C$.
In the following, we denote the Higgs branch by $\eta(C,D,\cA)$, where $D=\{D(\rho_i)\}$.
We also denote it as $\eta(C_{\rho_1,\ldots,\rho_n},\cA)$.

\item
It is worth remarking that the good/ugly/bad trichotomy can also be detected by studying the
virtual dimension of the mass-dimension $N$ part of the Coulomb branch of the would-be 4-dimensional field theory of the
$\cA \to 0$ limit, when $g=0$.
Each defect is characterized by $\rho_i$.
%
%
By Riemann-Roch, the
virtual dimension of the mass-dimension $N$ part of the Coulomb branch of $S_N[C,D]$ is
\begin{equation}
\dim_{\bC} \CM^\text{Coulomb}_{\text{mass dim}=N} = -(2N-1)+ \sum_{i} (N-\lambda_{i,1})
\end{equation}
Then the good/ugly/bad trichotomy corresponds to the cases where
$\dim_{\bC} \CM^\text{Coulomb}_{\text{mass dim}=N} $ is positive, zero, and negative, respectively.

\item In \cite{Gaiotto:2008ak} the good/bad/ugly trichotomy was established by
studying the conformal dimensions of monopole operators. In the good cases
the monopole operators have positive dimension, as computed from the R-symmetry.
In the the ugly cases, they are free fields. In  the bad cases they
have dimensions violating the unitarity bound as computed from the naive
R-symmetry. In the bad cases one  thus concludes that the IR R-symmetry must be different
from the UV R-symmetry.
These monopole operators come from  monopole strings in the 5d SYM theory wrapped on $C$.
 These in turn come from the surface defects of the 6d theory wrapped on $C$.
  Their holographic duals are then given by M2-branes wrapped on $C$.
In this last setting the associated chiral operators of the 4d theory
 were considered in  \cite{Gaiotto:2009gz}.
This is useful since,
in principle, one could compute the conformal dimensions of
these operators via the AdS/CFT correspondence.

\item Although we are focused here on the $\cA \to 0$ limit, it is worth noting
that the $\cA \to \infty$ limit is a weak coupling limit, and in this limit the
metric on the moduli space approaches a product metric on a fibration
over $(\bR^3 \times S^1)^{N-1}/\mathfrak{S}_N$ whose fiber is $\prod_i \cM^{\rm Coul}(T_{\rho_i})$.
Recall from \cite{Gaiotto:2008ak} that
$\cM^{\rm Coul}(T_{\rho_i}) = S_{\rho_i} \cap \cN$ is the intersection of a
Slodowy slice with the nilpotent cone. Thus, in the $\cA \to \infty$ limit
the Higgs moduli space can be made rather explicit.
\end{enumerate}

\subsection{As a hyperk\"ahler quotient}\label{subsec:asQuotient}

\begin{figure}
\[
\includegraphics[width=.9\textwidth]{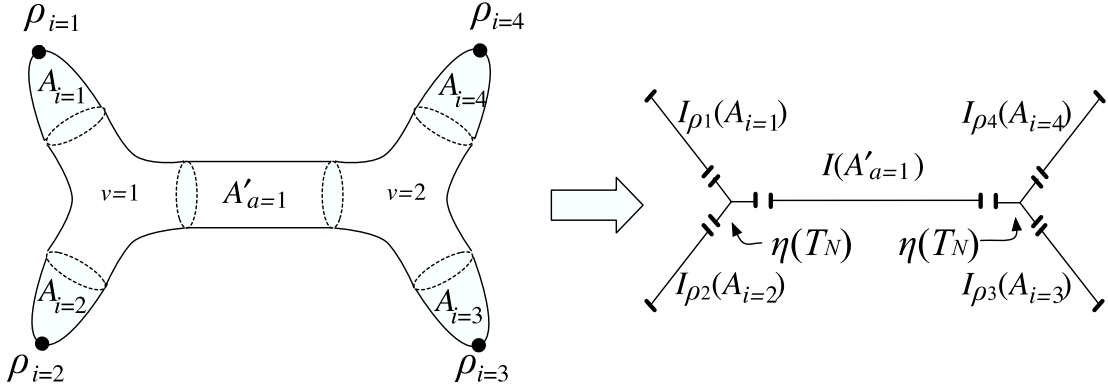}
\]
\caption{Left: a skeleton-like metric on $C$. Right: its reduction to 5d. The Higgs branch of the each component is named. The Higgs branch of the total system is given by the hyperk\"ahler quotient via the diagonal $\SU(N)$ actions.\label{skeleton}}
\end{figure}

As a second method, consider putting on $C$ a metric of cylinders of circumference $R_6$ joined at three-pronged junctures, so that the punctures are at the center of the caps, see Fig.~\ref{skeleton}.
The Higgs branch is then that of 5d maximally-supersymmetric Yang-Mills with coupling constant $1/g^2_{5d}\sim 1/R_6$ on a trivalent graph.
The original codimension-two defect at $p$ labeled by $\rho:\SU(2)\to\SU(N)$ becomes a supersymmetric boundary of the 5d Yang-Mills, given by \begin{equation}
\Phi_i(s)\sim \rho(t^i)/s + \Phi^0_i + {\cal O}(s)
\label{rhoBC}
\end{equation}
where $t^j = \frac{i}{4}\sigma^j$ is a basis of generators of
$\mathfrak{sl}(2,\IC)$,  $s$ is the distance to the boundary,
and   $\Phi^0_i$ must be in the commutant of $\rho$.
The junction of three segments is a supersymmetric boundary condition of $\SU(N)^3$ super Yang-Mills;
we have the 4d $T_N$ theory with $\SU(N)^3$ living on the boundary.
Therefore, the Higgs branch of this system is given by the hyperk\"aher quotient
 \begin{equation}
\eta(C,D,\cA)=\left[\prod_i I_{\rho_i}(\cA_i)_i  \times
\prod_{a} I(\cA_a')_{a,1;a,2} \times
\prod_{v} \eta(T_N)_{v,1;v,2;v,3} \right]
\hkq \prod_{j=1}^{3n_v} \SU(N)\label{HKQ}.
\end{equation}
Here, $I_\rho(\cA)$ is the moduli space of the Nahm equation on a segment of length $\cA$ with a boundary condition \eqref{rhoBC} on one side and with $\Phi_i(s)$ regular on the other side.
$I(\cA)$ is an abbreviation for $I_\rho(\cA)$ where $\rho$ is zero. Moreover
$\eta(T_N)$ is the Higgs branch of the 4d $T_N$ theory. The labels
$i$, $a$ and $v$ enumerate the external edges, the internal edges and the trivalent vertices respectively.
$\cA_i$ and $ \cA_a'$ are the areas of the external and internal
cylinders, respectively. The trinions carry zero area.

Here and in the following, we have actions of many copies of $\SU(N)$ on the spaces.
To distinguish them, we put subscripts to the spaces as in \eqref{HKQ},
so that  $\SU(N)_i$   act on $I_{\rho_i}(\cA_i)_i$,
$\SU(N)_{a,1}\times \SU(N)_{a,2}$ on $I(\cA'_a)_{a,1;a,2}$,
and $\SU(N)_{v,1}\times \SU(N)_{v,2}\times \SU(N)_{v,3}$ on $\eta(T_N)_{v,1;v,2;v,3}$.
The numerator of \eqref{HKQ} has an action by $6n_v$ copies of $SU(N)$. A subgroup, 
 defined by the   diagonal combinations of  $\SU(N)\times \SU(N)$ which are glued together, 
 and isomorphic to $3n_v$ copies of $SU(N)$, is gauged. In the following, we denote by $\SU(N)_{a,b}$ the diagonal subgroup of $\SU(N)_a \times \SU(N)_b$.

This construction is closely related to and partially overlaps with the bow construction of Cherkis and collaborators \cite{Cherkis:2008ip,Cherkis:2009jm,Cherkis:2010bn,Blair:2010kz,Blair:2010vh,Cherkis:2011ee}. Our $I_\rho(\cA)$ is their bow. Instead of their arrows, we have trivalent vertices.

$I_\rho(\cA)$ is a relatively well-studied manifold which we will review below;
the structure of $\eta(T_N)$ is also partially known.
Therefore this expression gives us a practical way to study the Higgs branch.
Note that this equality asserts that the hyperk\"ahler quotient on the right hand side only depends on
$\cA_i, \cA_a'$ through the sum $\cA=\sum \cA_i + \sum \cA_a'$. We will explain why this is so in Sec. \ref{subsec:Twistor}. First,
we  need to recall some basic properties of $I_\rho(\cA)$ and $\eta(T_N)$.

\subsection{The manifold $I_\rho(\cA)$}

An important special case of the above construction is the case where $C$ is a Riemann sphere with two punctures, where one puncture is characterized by $\rho$ and another puncture is full, $f=[1^N]$.
This gives the Higgs branch moduli space $I_{\rho}(\cA)$.
In the general notation this is  $\eta(C,\{D(\rho),D(f)\},\cA)=\eta(C_{\rho,f},\cA)$.

We review here the structure of the manifold $I_\rho(\cA)$, which is
 the moduli space of the Nahm's equation where $\Phi_i(s)$ satisfy the
 boundary condition \eqref{rhoBC} on one end,
 and are regular at the other end.
This moduli space was studied mathematically \cite{Kronheimer,Bielawski,Bielawski2} and was given physical interpretation in Sec.~3.9 of \cite{Gaiotto:2008sa}. We only quote salient results here; the details can be found op.~cit.

We already saw $I(\cA)$ in Sec.~\ref{T*G}.
As a holomorphic symplectic manifold, this is  $T^*\SU(N)_\bC$, which is further isomorphic to $\SU(N)_\bC \times \su(N)_\bC$  using the left-invariant one-forms.
The space $I_\rho(\cA)$ is, as a complex manifold, a subspace of $I(\cA)$ given by \begin{equation}
\SU(N)_\bC \times S_{\rho} \subset \SU(N)_\bC \times \su(N)
\end{equation} where $S_\rho$ is the Slodowy slice at $\rho(t^+)$, defined by \begin{equation}
S_{\rho}= \{ \rho(t^+)+x \bigm| [\rho(t^-),x] =0\}.
\end{equation}
 Here $t^\pm$ are raising/lowering operators in $\mathfrak{sl}(2)$.
 Note that the dimension of $S_{\rho}$ is the number of irreducible components of $\su(N)$ regarded as an $\SU(2)$ representation under the homomorphism $\rho$.
The complex moment map of the $\SU(N)$ action on $I_\rho(\cA)$ at $(g,X)\in \SU(N)_\bC\times S_\rho$ is $gXg^{-1}$.

From the description as the moduli space of the Nahm equation, it is clear that \begin{equation}
I_\rho(\cA+\cA')= I_\rho(\cA) \times I(\cA') \hkq \SU(N). \label{lengthening}
\end{equation}

As a side remark we note that, more generally, for the sphere with two \
punctures $D(\rho_1)$ and $D(\rho_2)$ the moduli space $\eta(C,\{D(\rho_1),D(\rho_2)\}, \cA)$
is the moduli space of solutions to Nahm's equations on the interval $[0,\cA]$
 with Nahm-type
boundary conditions of type $\rho_1, \rho_2$ at the two ends. As a
holomorphic manifold this is just
\begin{equation}
\{(g,X) \vert  X \in S_{\rho_1} \cap g^{-1} S_{\rho_2} g \} \subset T^*\SU(N)_\bC.
\end{equation}
A sphere with fewer punctures can be obtained by setting one or two of $\rho_{1,2}$ to be $[N]$, because a puncture with $\rho=[N]$ is equivalent to having no puncture. In particular, the sphere with no punctures at all corresponds to the manifold \begin{equation}
\{(g,X) \vert  X \in S_{[N]} \cap g^{-1} S_{[N]} g \} \subset T^*\SU(N)_\bC,
\end{equation} and in fact is the moduli space of centered BPS monopoles
with  $\SU(2)$ gauge group and magnetic charge $N$.\footnote{
The anomaly coefficients $n_v$ and $n_h$ of this theory from the sphere with no puncture can be calculated from the anomaly of 6d theory as was done for the good cases in pp. 19--21 of \cite{Benini:2009mz}. In the end we end up with putting $g=0$ in the universal formula (2.5) in \cite{Gaiotto:2009gz}, namely we have \[
n_v=-\left(\frac{4N^3}3-\frac{N}3-1\right),\qquad
n_h=-\left(\frac{4N^3}3-\frac{4N}3\right).
\] Note that they are negative, while in a superconformal theory both $n_h$ and $n_v$ are positive as shown in \cite{Hofman:2008ar,Shapere:2008zf}. This negativity of $n_h$ and $n_v$ also tells us that the theory is bad and that the $\cA\to 0$ limit cannot be easily taken. }

\subsection{The sphere with three full punctures and the manifold $\eta(T_N)$}

We now consider the case where $C$ is a sphere with three full punctures.
The 4d $\cA \to 0$ limit leads to the trinion theories $T_N$ introduced in
\cite{Gaiotto:2009we}.
We denote its  Higgs branch by $\eta(T_N)$.
In our general notation this is $\eta(C,\{D(f),D(f),D(f)\},0)=\eta(C_{fff},0)$.

The space $\eta(T_N)$ is  known to have the following properties \cite{Gaiotto:2009gz,Benini:2009mz,Benini:2010uu,Moore:2011ee}.
It is a hyperk\"ahler cone whose complex dimension is \begin{equation}
\dim_\bC \eta(T_N)= 3(N^2-1)-(N-1),
\end{equation} with a triholomorphic action of $\SU(N)^3$.
It is also the Coulomb branch of the star-shaped 3d quiver gauge theory in the limit where all gauge coupling constants are taken to be infinite, or equivalently the Coulomb branch of the 3d $\SU(N)$ theory coupled to three copies of $T[\SU(N)]$ theory in the infinite coupling limit, as we recalled in Sec.~\ref{asCoulomb}. We stress that for $\eta(T_N)$ we have already taken the $\cA \to 0$ limit, so that it does
not contribute to the overall area in equation \eqref{HKQ}.

$\eta(T_2)$ is a flat hyperk\"ahler manifold $\bH^4$ with $\SU(2)^3$ action, $\eta(T_3)$ is the minimal nilpotent orbit of $(E_6)_\bC$. $\eta(T_N)$ with $N>3$ is not explicitly known, but the following two important properties have been inferred from various dualities.

First, S-duality of two copies of the 4d $T_N$ theory coupled to $\SU(N)$, as described in \cite{Gaiotto:2009we}, implies the equality of the hyperk\"ahler manifold with action of $\SU(N)_1\times \SU(N)_2\times \SU(N)_3\times \SU(N)_4$:  \begin{equation}
\eta(T_N)_{1,2,a}
\times
\eta(T_N)_{3,4,b}
\hkq \SU(N)_{a,b}
=
\eta(T_N)_{1,3,a}
\times
\eta(T_N)_{2,4,b}
\hkq \SU(N)_{a,b}\label{TNcrossing}
\end{equation}
where $\eta(T_N)_{1,2,3}$ stands for $\eta(T_N)$ where $\SU(N)_1\times \SU(N)_2\times \SU(N)_3$ acts on it; the quotient is taken with respect to the diagonal subgroup of $\SU(N)_a\times \SU(N)_b$, which we denoted by $\SU(N)_{a,b}$.

A second important piece of information is about the moment maps. Let us denote the complex moment maps for $\SU(N)_i$ ($i=1,2,3$) as $(\mu_i)_\bC : \eta(T_N)\to \su(N)_\bC$. Then it is believed that $\tr (\mu_i)_\bC^k $ is independent of $i$. (See \cite{Benini:2010uu} for the argument.)
 In particular, \begin{equation}
\tr (\mu_1)_\bC^2 =\tr (\mu_2)_\bC^2=\tr (\mu_3)_\bC^2.\label{equalityOfQuadraticCasimirs}
\end{equation}

\subsection{Dependence on area from the perspective of the quotient}\label{subsec:Twistor}

\begin{figure}
\[
\includegraphics[width=.4\textwidth]{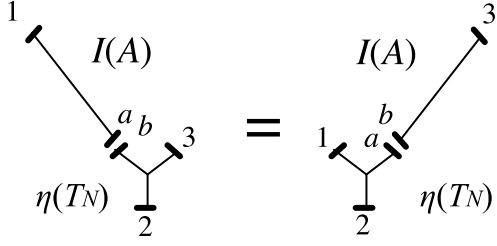}
\]
\caption{The property \eqref{moving} illustrated. The action of $\SU(N)_i$ is labeled by $i$ in the figure. \label{LR}}
\end{figure}

Readers interested mainly in the 4d theories can skip this and the next subsections and can directly go to Sec.~\ref{4d}.
Given \eqref{lengthening} and \eqref{TNcrossing}, the proof that the right hand side of \eqref{HKQ}  only depends on $\cA=\sum \cA_i + \sum \cA_i'$  and is furthermore independent of the pants decomposition boils down to the property \begin{equation}
I(\cA)_{1,a} \times \eta(T_N)_{b,2,3} \hkq \SU(N)_{a,b}
=
\eta(T_N)_{1,2,a} \times I(\cA)_{b,3} \hkq \SU(N)_{a,b},\label{moving}
\end{equation} see Fig.~\ref{LR}.

To show this, let us consider a more general procedure, which we can call the
\emph{hyperk\"ahler modification}.\footnote{This construction was introduced in Sec.~5 of \cite{Dancer:2010ke}; our small contribution is the explicit determination of the change in the twistor space and the hyperk\"ahler metric.}
 Let $Y$ be a hyperk\"ahler manifold with a triholomorphic action of $G$,
whose moment map is (after a choice of complex structure) $(\mu_\bC,\mu_\bR)$.
 We define the modification $Y(\cA)$ to be
\begin{equation}
Y(\cA)_1= I(\cA)_{1,a}\times Y_b \hkq G_{a,b}.
\end{equation}
 As a holomorphic symplectic manifold, $Y(\cA)$ is the same as the original $Y$: first, note that
\begin{equation}
T^*G_\bC \times Y \hkq G = \{ (g,X,y) \in G_\bC\times \fg_\bC \times Y\, |\, X+\mu_\bC(y)=0  \}  / \sim
\end{equation} where $(g,X,y)\sim (gh,h X h^{-1}, h\cdot y )$, where $h\cdot y$ stands for the action of $h\in G$ on $y\in Y$.
Then $X$ is determined by $\mu_\bC(y)$ and $g$ can be gauge fixed to be the identity.
Therefore, as a complex manifold, $Y(\cA)$ is canonically identified with $Y$. One can also check that the holomorphic symplectic form does not change.

The hyperk\"ahler metric, however, changes. The way it changes can be found by studying the twistor space,
 at least in the case  that $Y$ has an $SO(3)$ group of isometries which
rotate the three complex structures. In particular, this applies when $Y$ is a
Higgs branch of a four-dimensional $\cN=2$ theory, and also to the Higgs branch at finite $\cA$.
Recall that Hitchin's theorem
states that, roughly speaking, the twistor family of holomorphic symplectic manifolds
is equivalent to the \hk\ metric.

The twistor space of $I(\cA)$ was found by Kronheimer \cite{Kronheimer}:
 the transition function at the equator is given by \begin{equation}
(g,X,\zeta)\to (g \exp(2\cA X/\zeta), X \zeta^{-2},\zeta')
\end{equation} where $(g,X)\in G_\bC\times \fg_\bC$ and $\zeta'=1/\zeta$.
Using the $SO(3)$ isometry rotating the three complex
structures,  the twistor space of  $Y$ can be holomorphically trivialized
on the northern and southern hemispheres of the twistor sphere and hence
the twistor space of $Y$ can be presented as
 \begin{equation}
(y,\zeta)\to (\phi_\zeta(y),\zeta').
\end{equation}
Now, because the $G$-action is triholomorphic the
moment map satisfies
\begin{equation}
\zeta^{-2} \mu_{\bC,\zeta}(y) = \mu_{\bC,\zeta'}(\phi_{\zeta}(y))
\end{equation}
and hence the equation
$X +  \mu_{\bC,\zeta}(y) =0$ is consistent across patches.
After choosing the gauge $g=1$ and eliminating $X$ we find that
 the twistor space of $Y(\cA)$ is given by
 \begin{equation}
(y,\zeta)\to (\exp\left(- 2\cA \mu_{\bC,\zeta}(y)/\zeta\right) \phi_{\zeta}(y),\zeta').\label{twistormod}
\end{equation}
Infinitesimally, the action of $\exp(2\cA \mu_{\bC,\zeta}(y)/\zeta)$ on $Y$ is generated
by the vector field $v= 2\sum_a \mu_{\bC,\zeta}^a v^a /\zeta$ where $a$ is the adjoint index and $v^a$ is the vector field for the $a$-th generator of $G$.
It is easy to see that $v$ is in fact the Hamiltonian vector field for $\tr \mu_{\bC,\zeta}^2/\zeta$.
Therefore, the deformation $Y(\cA)$ of $Y$ is determined once the quadratic Casimir of the complex moment map, $\tr \mu_{\bC,\zeta}^2$ is given for every complex structure $\zeta$. Applying this constructing to $\eta(T_N)$ and
invoking the property \eqref{equalityOfQuadraticCasimirs} we establish   \eqref{moving}, the   desired identity.

In fact, we can say a little more about how the metric is deformed by $\cA$.
Using the results of \cite{Alexandrov:2008ds} (see their eq.~(4.27)), we can extract the modification of the K\"ahler potential from the modification \eqref{twistormod} in the twistor construction. Denoting $K(\cA)$ be the K\"ahler potential of $Y(\cA)$ in the complex structure $I$, we have \begin{equation}
\frac{d}{d\cA} K(\cA) = \mathrm{Re}\ \tr\ \mu_\bC^2
\end{equation} where $\mu_\bC$ is the complex moment map in the complex structure $J$.

\subsection{A connection with the bow construction}
Before proceeding, let us consider $\eta(C,\{D(f),D(f),D(s)\},\cA)$ for a three-punctured sphere $C$.
At $\cA=0$, this is just the bifundamental hypermultiplet Higgs branch $B_{1,2}=\bC^{N^2}\oplus \bC^{N^2}$ with a natural $\SU(N)_1\times \SU(N)_2$ action.
Then at nonzero $\cA$, it is given by \begin{equation}
I(\cA')_{1,a}\times B_{b,c} \times I(\cA'')_{d,2} \hkq \SU(N)_{a,b}\times \SU(N)_{c,d}
\end{equation}  where $\cA=\cA'+\cA''$.
That this quotient only depends on the sum $\cA'+\cA''$ follows from the fact that $\tr \mu_\bC^{(1)}{}^2=\tr (AB)^2$ and $\tr \mu_\bC^{(2)}{}^2=\tr (BA)^2$ are equal.
The dependence only on the sum is also known in the context of the bow construction. As shown in \cite{Blair:2010vh}, this is a stratum in the moduli space of $\SU(2N)$ BPS monopoles on $\bR^3$ with one Dirac singularity where the vev of the adjoint scalar is given by $\diag(\underbrace{a,\ldots,a}_N,\underbrace{-a,\ldots,-a}_N)$. The difference $\cA'-\cA''$ gives the B-field on $\bR^3$ but it does not affect the moduli metric.

\section{Application to the 4d analysis}\label{4d}

We now return to the subtleties in the factorization statement
\eqref{eq:glue-gauge}. When $C$ factorizes we expect the Higgs
branches of the theories to be related by \hk\ gluing. In particular,
if we factorize on full punctures then the diagonal of the global $\SU(N)\times \SU(N)$
symmetry of the $C_L \sqcup C_R$ theory is gauged. Thus, the moment
maps $\mu_L$ and $\mu_R$ of the flavor symmetries at $p_L, p_R$
are identified: $\mu_L + \mu_R =0$. Now suppose that the six-dimensional
theory associated to $C_R$ is bad.   Then,  no point on
the Higgs branch has $\mu_R=0$. Then the $\SU(N)$ symmetry is always
spontaneously broken, and the $\cA \to 0 $ limit forces $\mu_R$,
and hence $\mu_L $ to go to infinity. The vacuum flows to that of
a new theory, and in the limit the gauge symmetry can become smaller.
In the ``ugly'' theories, there is a point where $\mu_R=0$.
We now examine some special cases of such bad and ugly theories
by studying some trinion theories $C$ with defects $D(\rho_1), D(\rho_2), D(\rho_3)$.
We denote them by $C_{\rho_1,\rho_2,\rho_3}$.

\subsection{$T_N$ and the bifundamental}

First, let us compare a sphere $C_{fff}$ with three full punctures and a sphere $C_{ffs}$ with two full punctures and one
 simple puncture. Recall that the full puncture is $f=[1^N]$ and the simple puncture is $s=[N-1,1]$.
At finite non-zero area, the Higgs branch of $C_{ffs}$ is given by \begin{equation}
\eta(C_{ffs},\cA)_{1,2}=I_s(\cA)_a\times \eta(T_N)_{b,1,2} \hkq \SU(N)_{a,b}. \label{foo}
\end{equation}
As a complex manifold, $I_s(\cA) \simeq \SU(N)_\bC \times S_s$ as discussed before.
 As $\dim_\bC S_s=N+1$, $\dim_\bC I_s(\cA)=N^2+N$.
The $\SU(N)$ action  on $I_s(\cA)$ is free. Therefore the dimension of $\eta(C_{ffs},\cA)$ is \begin{equation}
\dim_\bC \eta(C_{ffs},\cA)=\dim_\bC \eta(T_N)+\dim_\bC I_s(\cA)- 2\dim \SU(N) = 2N^2.
\end{equation}
If we use the mirror quiver as in Sec.~\ref{asCoulomb}, the central node has $N_f=2N-1$, and is ``ugly,''
in the terminology of \cite{Gaiotto:2008ak}.  Therefore we expect to have $N^2$ free hypermultiplets in the $\cA\to 0$ limit.
This matches our expectation that the 6d theory on $C_{ffs}$ at zero area gives the bifundamental hypermultiplet of $\SU(N)^2$.
For $N=3$, the equation of $\eta(T_3)$ is known \cite{Gaiotto:2008nz}, and the quotient \eqref{foo} can in principle be explicitly performed at the level of the holomorphic symplectic quotient.

\subsection{The trinion  with one full and two simple punctures}\label{fss}

Next, let us consider a sphere $C_{fss}$ with one full puncture and two simple punctures.
At finite nonzero area, the Higgs branch is given by \begin{equation}
\eta(C_{fss},\cA)_{1}=I_s(\cA)_a\times \eta(C_{ffs},0)_{b,1} \hkq \SU(N)_{a,b} \label{bar}
\end{equation} where $\eta(C_{ffs},0)$ is the linear space of a bifundamental.
The dimension is easily calculated: 
\begin{equation}\label{eq:fss}
\dim_\bC \eta(C_{fss},\cA)=\dim_\bC \eta(C_{ffs},0)+\dim_\bC I_s(\cA)- 2\dim \SU(N) = N^2+N+2.
\end{equation}
This space has a triholomorphic action of $\SU(N)$, but there is no point on this
 space where it is unbroken: if it were unbroken then the dimension
 would need to be at least $2(N^2-1)$. But $2(N^2-1)$ is larger than \eqref{eq:fss}
 for $N>2$.

We believe that there is another equivalence of the hyperk\"ahler spaces \begin{equation}
\eta(C_{fss},\cA)_1= I_t(\cA)_1 \times (\bC^2\oplus \bC^2) \hkq \SU(2)\label{zot}
\end{equation} where $t$ is the partition $[N-2,1,1]$, and the $\SU(2)$ action is the diagonal action between the commutant of $t(\SU(2))$ inside $\SU(N)$ and a natural action of $\SU(2)$ on $\bC^2\oplus \bC^2$.
This equality can in principle be proven by expressing the right hand sides of \eqref{bar} and \eqref{zot} as the moduli spaces of the Nahm equation. This relation can also be inferred from the analysis of the S-dual of $\SU(N)$ with $2N$ flavors \cite{Chacaltana:2010ks}.

\begin{figure}
\[
\includegraphics[width=.7\textwidth]{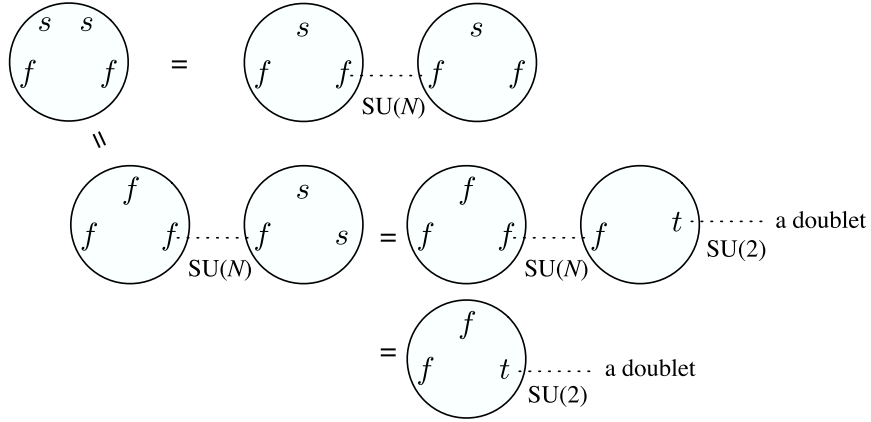}
\]
\caption{A sphere with punctures $f,f,s,s$ and two ways of decomposition. Every step is understood to be performed at finite area. \label{ffss}}
\end{figure}

Now we can have a new insight why we had $\SU(2)$ as the gauge symmetry in the strong-coupling limit of $\SU(N)$ theory with $2N$ flavors. As in \cite{Gaiotto:2009we}
we start from a sphere with two full punctures and two simple punctures, $C_{ffss}$.
When two simple punctures are close, at finite nonzero area $\cA$,
we have a  sphere $C_{fff}$ coupled to a sphere $C_{fss}$ with area $\cA$.
Now the latter is equivalent to a two-punctured sphere $C_{ft}$ with area $\cA$ coupled to a doublet $\bC^2\oplus \bC^2$ of $\SU(2)$.
So, we have $C_{fft}$ with area $\cA$ coupled to a doublet of $\SU(2)$ by a gauge group $\SU(2)$, see Fig.~\ref{ffss}. At this final stage we can safely take the $\cA\to0$ limit.

When $N=3$ the analysis can be stated more simply, since
 the puncture $t$ is the full puncture $f$. In this case, $\dim_\bC \eta(C_{fss},\cA)=14$. The $\SU(3)$ action is broken to $\SU(2)$. Ten of the chiral multiplets give mass to the $5$ broken generators, leaving four chiral multiplets charged under $\SU(2)$, which is in fact in the doublet.
So, we have $C_{fff}$ coupled to $C_{fss}$, via $\SU(3)$ gauge group. But this is spontaneously broken to $\SU(2)$ because of the property of $C_{fss}$, leaving a doublet of  $\SU(2)$ in the $\cA \to 0$ limit.

\subsection{A sphere with four punctures of type $[k,k]$}\label{sec:kk}

As a final example, let $N=2k$ and consider a sphere with four punctures of type $[k,k]$.  This is dual to $k$ D3-branes probing a $D_4$-type singularity, i.e. an orientifold 7-plane with four D7-branes on top of it.
Therefore the 4d field content is $\Sp(k)$ with four fundamentals and one antisymmetric.
\footnote{By $Sp(k)$ we mean the compact group of real dimension $2k^2+k$. In particular, $\Sp(1)=\SU(2)$.}

\begin{figure}
\[
\includegraphics[width=.9\textwidth]{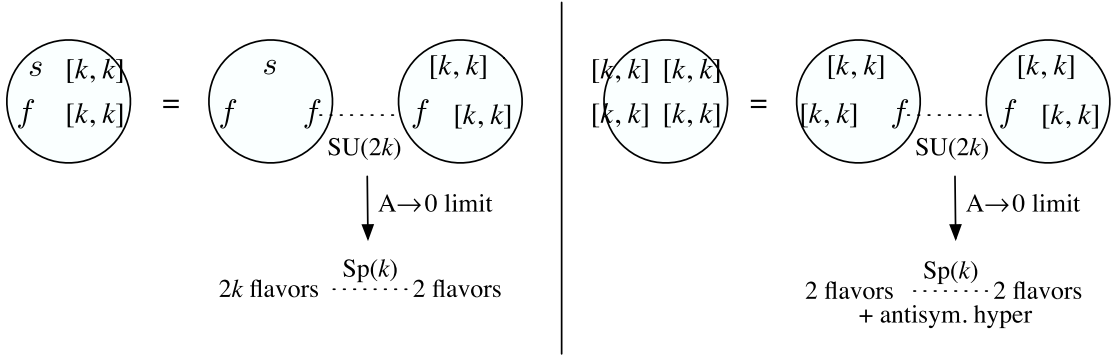}
\]
\caption{A sphere with punctures $f,s,[k,k],[k,k]$, and a sphere with four punctures of type $[k,k]$. The appearance of an additional antisymmetric in the second case can be understood naturally in our approach \label{kk}. }
\end{figure}

In \cite{Gaiotto:2009we,Nanopoulos:2009xe,Chacaltana:2010ks}, it was noted that a  sphere with four punctures $f$, $s$ and two $[k,k]$'s realizes  $\Sp(k)$ theory with $2k+2$ flavors,
which is perturbatively conformal.
By splitting the sphere,  $2k$ flavors can be accounted for as coming from $C_{ffs}$, which
gives $2k$ flavors of $\SU(2k)$. Then two punctures of type $[k,k]$ were thought of as somehow restricting the gauge group to be $\Sp(k)$, and
 moreover providing $2$ more flavors, see the left side of Fig.~\ref{kk}.

We interpret this as saying that the Higgs branch $X=\eta(C_{[k,k],[k,k],f})$ has
an action of  $\SU(2k)$ on the full puncture, but it is  always broken to $\Sp(k)$.
At a point where $\Sp(k)$ is preserved, one has two fundamentals.
Then, if we glue this to $C_{ffs}$, the $\SU(2k)$  is spontaneously broken to $\Sp(k)$,
and  the directions of $X$ representing the broken directions in $\SU(2k)/\Sp(k)$ are eaten by the massive vector bosons.

Now, consider the sphere with four punctures of type $[k,k]$. This is obtained by gluing two copies of $C_{[k,k],[k,k],f}$ at the full puncture.
Now, there is only one $\SU(2k)$ gauge group, which is broken to one $\Sp(k)$.
However, we have two copies of manifold $X$, with two copies of broken directions  $\SU(2k)/\Sp(k)$.
Only one of them is eaten by the Higgs mechanism, and one remains as a physical direction,  transforming as an antisymmetric of $\Sp(k)$.
By taking the $\cA\to 0$ limit, one finds $\Sp(k)$ theory with four flavors and an antisymmetric, as expected
from the orientifold picture.
In comparison, in the approach of \cite{Chacaltana:2010ks}, the appearance of the antisymmetric needs to be put in by hand.

\section{OPE of  Codimension-Two Defects}\label{ope}

In this section we would like to use some of the lessons learned
from the limits and compactifications we have studied to learn
 about the six-dimensional $(2,0)$ theory itself in $\bR^{1,5}$. Namely,
we can consider the behavior in $\bR^{1,5}$ when
two half-BPS codimension-two defects $D(\rho)$ are placed parallel
to each other and are brought together.
Suppose the transverse plane is
identified with $\bC$ and one defect sits at $z=0$ while the other is at
a point $z$. What happens as $z\to 0$ ?

In order to answer this question it is necessary to enlarge the set of
half-BPS defects under consideration. When   $D(\rho)$ has global symmetry $H$ it can
be coupled to any 4d N=2 field theory
with $H$-global symmetry, say $T^4_H$, by gauging the diagonal
global symmetry, as in \eqref{eq:glue-gauge}. This gives a new defect
\begin{align}\label{eq:glue-gauge-one}
D(\rho, T^4_H, q) &:= D(\rho) \times_{H,q} T^4_H \\
&= \int [d\Phi] e^{i 2\pi \int d^4 x d^4 \theta \tau \tr\Phi^2/2+c.c.}
T^4_H(\Phi)  D(\rho; \Phi) \label{eq:pathintegral}
\end{align}
where $\Phi$ stands for the $\cN=2$ vector superfield;
$D(\rho;\Phi)$  and $T^4_H(\Phi)$ stand for the defect $D(\rho)$ and the theory $T^4_H$ coupled to the external vector superfield $\Phi$.
In   particular the lowest component of $\Phi$ serves as mass parameters for $D(\rho)$ and $T^4_H$.
The path integral \eqref{eq:pathintegral} then makes $\Phi$ dynamical, with coupling constant $\tau$.

To make this path integral UV complete,
 there is a bound on the flavor  central charge $k$ of the global $H$ currents of the two components.
Namely,
\begin{equation}
k(T^4_H) + k(D(\rho)) \le k(\text{adjoint hyper of $H$}).
\end{equation} $k$ is proprotional to the contribution of the theory to the one-loop beta function of $H$ gauge fields
\cite{Erdmenger:1996yc,Anselmi:1997rd,Argyres:2007cn}.
When this equality is not saturated $e^{i\pi\tau}$ undergoes dimensional
transmutation to   a dimensionfull scale.
Then, defects preserving the conformal invariance should saturate the bound
if we want a 4d superconformal theory.

Suppose we have $D(\rho_1)$ at $z=0$ and $D(\rho_2)$ at $z$.
Then, from far away, there should be an effective defect representing the two.
We conjecture that it is
of the type $D(\rho_3, T^4_H, q)$  where $q= z = e^{i \pi \tau}$.
The precise rules for determining $\rho_3$ from $\rho_1, \rho_2$ can be extracted
from Section 4.5  of \cite{Gaiotto:2009we} and from \cite{Chacaltana:2010ks}. We will see a few examples momentarily.

We can express this operation as
\begin{equation}
\begin{split}
D(\rho_1)_z D(\rho_2)_0&\sim D(\rho_3, T^4_H, z) \\
 &=  \int [d\Phi] z^{S[\Phi]} \bar z^{S[\bar\Phi]} T^4_H(\Phi) D(\rho_3;\Phi)
\end{split}
\end{equation} where $S[\Phi]=2\pi \int d^4 x d^4 \theta \tr\Phi^2/2$.

Comparing this to the standard OPE $ \cO_1(z,\bar z) \cO_2(0)\sim \sum_i z^{\Delta_i} \bar z^{\bar\Delta_i }c^i_{12} O_i(0) $,
we see that  the
conformal dimensions $\Delta_i$ are formally
reinterpreted as the action of the vector multiplet $\Phi$, while the
four-dimensional   quantum field theory $T^4_H$ appears
as an ``operator product expansion coefficient.''
Furthermore, it is a path integral, instead of a summation.
The appearance of
a four-dimensional field theory as an operator product expansion
coefficient generalizes the vector-space-valued OPE coefficients
of line defects discussed in \cite{Gaiotto:2010be}. We expect this
idea will fit in naturally with the general ideas of extended
topological field theories currently under development by
several physical mathematicians.

For examples, we can rewrite what we learned in Sec.~\ref{4d} in the language of the OPE.
The analysis of $C_{fss}$ leads us to the equality \eqref{zot} in Sec.~\ref{fss}. This can be though of as the OPE \begin{equation}
D(s)_z D(s)_0 \sim  D(t,(\text{one flavor})_{\SU(2)},z),
\end{equation} where $(\text{one flavor})_{\SU(2)}$ stands for the theory  of free hypermultiplets in the doublet of $\SU(2)$.
 Similarly the analysis of $C_{f,[k,k],[k,k]}$ in Sec.~\ref{sec:kk} tells us that \begin{equation}
D([k,k])_z D([k,k])_0 \sim D(f, (\text{two flavors})_{\Sp(k)},z).\label{xxx}
\end{equation}

In this language, the sphere with four punctures of type $[k,k]$ can be analyzed as follows.
First, we take the OPE of the two pairs using \eqref{xxx}. Then, we have a sphere with two defects of type $D(f, (\text{two flavors})_{\Sp(k)},z)=D(f)\times_{\Sp(k)} (\text{two flavors})$.
Equivalently, we have a sphere with two full punctures,  each coupled to two flavors via $\Sp(k)$. The sphere with two full punctures produces a theory with Higgs branch $I(\cA)=T^*\SU(2k)_\bC$. We are gauging this theory via $\Sp(k)^2$ from the left and the right simultaneously. This breaks $\Sp(k)^2$ to $\Sp(k)$, and a part of $I(\cA)$ remains as the antisymmetric of $\Sp(k)$.
Taking $\cA\to 0$ limit, we have 4d $\Sp(k)$ theory coupled to four flavors plus an antisymmetric.


\section*{Acknowledgements}
The authors thank Sergey Cherkis,  Jacques Distler, Andy Neitzke, and
Edward Witten for discussions.
The work of DG is supported in part by NSF PHY-0969448 and also by
the Roger Dashen Membership.
The work of GM is supported by the DOE under grant
DE-FG02-96ER40959.
The  work of YT is  supported in part by World Premier International Research Center Initiative
(WPI Initiative),  MEXT, Japan through the Institute for the Physics and Mathematics
of the Universe, the University of Tokyo.

\bibliographystyle{ytphys}
\small\baselineskip=.95\baselineskip
\bibliography{ref}

\end{document}